\documentclass[conference]{IEEEtran}

\usepackage[linesnumbered,lined, algoruled]{algorithm2e}
\usepackage{amssymb,bm}
\usepackage{amsmath}
\usepackage[acronym,shortcuts]{glossaries}
\usepackage{algorithmic,float}
\usepackage{balance}
\usepackage{color}
\usepackage{dsfont}
\usepackage{soul}
\usepackage{graphicx}
\usepackage{siunitx}
\usepackage{comment}
\newtheorem{theorem}{Theorem}
\newtheorem{lemma}[theorem]{Lemma}
\newtheorem{proposition}[theorem]{Proposition}

\begin{document}
\title{Blind Coherent Preamble Detection via Neural Networks}

\author{Jafar Mohammadi, Gerhard Schreiber, Thorsten Wild, Yejian Chen\\
Nokia Bell Labs, Stuttgart, Germany}

\maketitle

\sloppy

\begin{abstract}
In wireless communications systems, the user equipment (UE) transmits a random access preamble sequence to the base station (BS) to be detected and synchronized. In standardized cellular communications systems Zadoff-Chu sequences has been proposed due to their constant amplitude zero autocorrelation (CAZAC) properties. The conventional approach is to use matched filters to detect the sequence. Sequences arrived from different antennas and time instances are summed up to reduce the noise variance. Since the knowledge of the channel is unknown at this stage, a coherent combining scheme would be very difficult to implement. 

In this work, we leverage the system design knowledge and propose a neural network (NN) sequence detector and timing advanced estimator. We do not replace the whole process of preamble detection by a NN. Instead, we propose to use NN only for \textit{blind} coherent combining of the signals in the detector to compensate for the channel effect, thus maximize the signal to noise ratio. We have further reduced the problem's complexity using Kronecker approximation model for channel covariance matrices, thereby, reducing the size of required NN.
The analysis on timing advanced estimation and sequences detection has been performed and compared with the matched filter baseline.   
\end{abstract}


\section{Introduction}\label{sec:intro}
The competence of Neural Networks (NN) as a universal function approximation tool has been proved over and over for many use cases in wireless communications \cite{Debbah19, Air_HarishJacob21}. Recently, many results have emerged targeting the learning of even complex functions such as an entire receiver chain \cite{End2EndJakob19,Dani21}. The signal processing operations needed to build a full receiver has been also modeled by using NN, especially convolutional neural networks (CNN). Among them, we can name belief propagation decoding \cite{decoding18,Stb_decoding}, channel estimation and prediction \cite{MMSE2018, Schotten19, arnold2019}, and preamble sequence detection \cite{Chiara19}. The authors in \cite{TurboAI_1, MMSE2018} trained an NN to mimic the behavior of the minimum mean squared error estimator (MMSE), without the need for highly task of complex matrix inversion, which comes with the linear MMSE.  

The problem of preamble detection appears in the random access procedure, where the base station (BS) may not have any information on presence of a user equipment (UE). The UE sends the BS a sequence, which is randomly chosen from a set of sequences configured by the BS. The 3GPP\footnote{3rd Generation Partnership Project} standardized sequences are taken from Zadoff-Chu sequences with prime sequence length. The detection of this sequence from noise could be very tough, especially when the UE is at the cell-edge, i.e.\ very low signal to noise ratio (SNR) scenarios. This problem is formally presented as a classical signal detection problem and solved by hypothesis testing tools. The optimal detector in the presence of additive noise and single component channel is the energy detector. However, in real world scenarios, the channel may have multipath components and this could further deteriorate the detection probability. In \cite{Chiara19}, the authors reported a huge sensitivity to the input signal SNR, thus they proposed training a neural network for each input SNR separately. This leads to a huge model trained and maintained at the BS, which could be very expensive.

In this work, we focus on the spatial (antenna) and temporal processing parts of the problem. We propose a scheme that can learn the coherent combinations of the sequences coming from different antennas and time instances and combine them to increase the SNR. To the best of our knowledge, coherent signal processing for preamble detection in random access channel has not been introduced. We reformulate the problem such that we could use the results on the well studied problem of channel estimation.  
We design an NN inspired by the work of \cite{MMSE2018} and \cite{TurboAI_2} to coherently combine the inputs from various antennas. We show that these two problems are mathematically similar, therefore the tools and experience is transferable.

In section II, we first introduce the problem statement and our system model. We present the model as general as possible, however to carry the message of this work better, we focus on a smaller setting initially. Section III presents the main idea of this work. We first focus on the two dimensional problem, i.e. frequency and spatial, for sake of ease of presentation, then we briefly present how to extend the results to higher dimension, i.e. frequency, spatial, and temporal. We conclude this work by presenting extensive numerical results on 3GPP compliant link level simulator in the section IV.  

\section{Problem statement}
In a standardize wireless cellular communications system, the preamble sequence is chosen from a predefined set of Zadoff-Chu sequences, where a Zadoff-Chu sequence could be constructed as: 
\begin{align*}
z_u[n] = e^{-j\pi u \frac{n(n+1)}{N_z}} , \;\; n \in \mathbb{Z} :n \in [0, N_z-1],
\end{align*}
where $u$ denotes the root sequence ID and $N_z$ is the sequence length.
Zadoff-Chu sequences are among constant amplitude zero autocorrelation (CAZAC) sequences. The cyclically shifted versions of Zadoff-Chu sequences, i.e. $z_{u,l} = z_u[(n+c_l)\; \text{mod} \;N_z]$, where $c_l$ is the cyclical shifts which is determined by the cell size \cite{Gerhard18} have zero correlation zones. The sequences are picked up by the BS with a time delay $d$, which is denoted by $\mathbf{Z}_{u,l,d} = \textbf{Diag}[z_{u,l,d}]$, where $\textbf{Diag}[\cdot]$ is an operator that takes a vector size $n$ as input and outputs a diagonal matrix size $n \times n$, with diagonal elements taken from the input vector.

Assuming a BS with $1\leq M$ receive antennas, the matrix carrying a preamble that occupies $1\leq S$ subcarriers is denoted by
\begin{align*}
\mathbf{X} = \mathbf{H} \mathbf{Z}_{u,l,d} + \mathbf{N},
\end{align*} 
where, $\mathbf{X},\mathbf{H}, \; \text{and} \; \mathbf{N} \in \mathbb{C}^{M \times S}$ are the received signal, the channel, and the noise observation, respectively. We further assume $E[\mathbf{N} \mathbf{N}^H ] = \sigma^2 \mathbf{I}_N$. Let's assume that UE selects $l$th cyclic shift of the root sequence $u$ at random, which has arrived with the time delay of $d$ and denoted by $\mathbf{Z}_{u,l,d} =\mathbb{C}^{S \times S} $.  
The conventional approach for this problem is matched filtering \cite{matchedFilter60}, i.e. first correlate the received signal with the root sequence for the cell, i.e. $\mathbf{Z}_u$ as: 
\begin{align*}
\mathbf{Y} = \mathbf{X} \mathbf{Z}^*_u =  \mathbf{H} \mathbf{Z}_{u,l,d}\mathbf{Z}^*_u + \mathbf{N} \mathbf{Z}^*_u,
\end{align*} 
and then forming a hypothesis testing. The energy detector would be the optimal solution in case of no prior knowledge: 
\begin{align*}
\sum_{s,m} |y[s,m]|^2 \gtreqless \gamma (P_f), 
\end{align*} 
where, $y_{s,m} \in \mathbb{C}$ is the $(s,m)$th element of $\mathbf{Y}$, $\gamma$ is the threshold according to a given false alarm probability $P_f$ \cite{JafarJamming15}. If we are given the channel values, we could provide a weighted energy detector, i.e. 
\begin{align*}
\sum_{s,m} |w[s,m] \;y[s,m] |^2 \gtreqless \gamma (P_f), 
\end{align*}
where, the $w[s,m] \in \mathbb{C}$ is the coherent coefficients for spatial and frequency combining. 

From system design viewpoint it is impractical to assume the knowledge of the channel for preamble detection, since the preamble signal is emitted at the initial access. Therefore, finding the optimal weights $w[s,m]$ might not be feasible. In the next session, we propose a solution to tackle this problem based on neural networks. Note that the model considered above we have considered temporal repetition of the sequences for the sake of simplicity of the notation. In the end of the next session we extend our solution to consider also temporal dimension.  

\section{Blind Coherent Antenna Combining}

To find the coherent coefficients $w[s,m]$, we propose to use a data driven estimation method based on NN. Basically the NN learns the statistical characteristics of the channel. For a coherent combining receive scheme, we require the knowledge of the channel at each antenna, subcarrier and time. 
We propose to learn this by NN trained on large dataset. In a supervised learning scheme, we can search for different NN architectures and perform Baysian optimization to find the best hyperparameters, e.g. NN layers type and number, non-linearity type, optimizer and its parameters, etc. 
However, there are a few issues with such an approach. Firstly, due to the large size of the input tensors e.g.\ 53696 for 64 antennas and 839 subcarriers, the required NN size and therefore the computational complexity skyrockets. The second challenge of this approach is the training procedure. To train such a large NN, one has to also consider that the sample complexity of the NN increases, i.e. the required number of samples for training to achieve an arbitrary accuracy.
In the rest of this section, we propose methods to decrease complexity of such a NN, by exploiting the underlying statistical properties of the preamble sequences. We further use the inherit redundancy in the received matrix (tensor) to reduce the size of the required NN. 

\subsection{Breaking the curse of dimensionality}
The input tensor has two dimensions that could be decomposed into two subproblems by the Kronecker model for channel covariance matrices approximation \cite{Veeravalli03,Kronecker}. This decomposition is a good approximation to the second order statistics of the larger tensor. We can decompose the tensor into two smaller tensors and tailor a neural network solution for each of these smaller problems rather than one gigantic problem \cite{TurboAI_2}.

Formally, the Kronecker decomposition approximation for the channel covariance matrix is given as: 
\begin{align} \label{eq:knonecker}
E[\mathbf{GG}^{H}] \approx E[\mathbf{h}_f \mathbf{Z}_{u,d,l} \mathbf{Z}^*_u (\mathbf{h}_f \mathbf{Z}_{u,d,l} \mathbf{Z}^*_u)^H] \otimes E[\mathbf{h}_a\mathbf{h}_a^H]]
\end{align}
where, $\mathbf{G} \in \mathbb{C}^{SM \times 1}$ is the vectorized form of the channel tensor and Zadoff-Chu sequence after the matched filter, $ \mathbf{h}_f\in \mathbb{C}^{S \times 1}$ and $\mathbf{h}_a\in \mathbb{C}^{M \times 1}$ are channel values (observations) for frequency and antenna directions only. Note that the size of the two problems combined is still smaller than one large problem.  

One further step toward reducing the dimensionality of the problem is to consider the next step, i.e. building the MMSE estimator matrix. 
\begin{proposition}
Assuming the Kronecker decomposition approximation holds tightly in (\ref{eq:knonecker}), 
then the MMSE on the frequency direction, 
\begin{align*}
\mathbf{R}_f:=E[\mathbf{h}_f \mathbf{Z}_{u,d,l} \mathbf{Z}^*_u (\mathbf{h}_f \mathbf{Z}_{u,d,l} \mathbf{Z}^*_u)^H],
\end{align*}
does not improve the overall estimation accuracy, due to the CAZAC properties of the Zadoff-Chu sequence.  
\end{proposition}

\textit{Sketch of the proof:}
The sub-problem in the frequency direction could be re-written as:
\begin{align*}
\min_{\mathbf{V}} \|\mathbf{V} \mathbf{y}_m - \mathbf{h}_f\|_2^2, \;\; \forall m 
\end{align*}
where, $\mathbf{y}_m = \big[y[0,m], y[1,m], \dots, y[S,m]\big]^H$ for any arbitrary $m$ and $\mathbf{V} \in \mathbb{C}^{S \times S}$ is the weight matrix. This is the MMSE problem, for which the optimal solution is given as:
\begin{align}\label{eq:mmse}
\mathbf{V} = \mathbf{R}_f (\mathbf{R}_f+ \sigma^2 \mathbf{I}_N )^{-1}
\end{align} 
where, $\mathbf{R}_f:= E[ \mathbf{h}_f\mathbf{Z} (\mathbf{h}_f\mathbf{Z})^H] $ is the covariance matrix of the channel after the matched filter. To proceed with the proof, we use the zero autocorrelation property as it is given in the following lemma:
\begin{lemma}[\cite{popovic20}]
The periodic autocorrelation function of the Zadoff-Chu sequence with $N_z$ and $u$ being coprime, is zero for cyclical shifts $n<N_z$. The same holds for its Discrete Fourier Transform (DFT). 
\begin{align}
\sum_{l=0}^{N_z-1} z_u[n]z^*_{u,l}[n] =
\sum_{l=0}^{N_z-1} z_u[n]z^*_u[(n+c_l) \; \text{mod} \;N_z] = 0.
\end{align}
\end{lemma}
According to the above lemma, and the fact the distribution of the arrival sequences $Z^*_{u,l}$ and the delay $d$ in a given cell is uniform over all the cyclic shifts $l$ and $d$, we can conclude that $\mathbf{R}_f$ is diagonal. In other words, the CAZAC properties of the Zadoff-Chu whitens the correlation among the neighboring subcarriers. Note that this phenomena does not occur in the spatial domain as all the BS antennas receive the same transmitted signal.  

For a diagonal covariance matrix, the MMSE simplifies to the least squared (LS) estimator, which is just a simple coefficient multiplication. Therefore, dedicating a NN to learn the MMSE estimator in the frequency direction does not improve the overall performance. The following lemma captures this conjecture in a more formal way: 
\begin{lemma}\label{lem:CovMatMMSE} 
Consider two Hermitian positive semi-definite matrices $\mathbf{A}$ and $\mathbf{B}$ describing two different covariance matrices for $\mathbf{h}$ in $\mathbf{y} = \mathbf{h} + \mathbf{n}\; \in \mathbb{C}^{M \times 1}$ with only difference in off-diagonal values and:  
\begin{align*}
\|\mathbf{A}\|_F \leq \|\mathbf{B}\|_F. 
\end{align*}
Further consider the MMSE resulting from $\mathbf{A}$ and $\mathbf{B}$, $\mathbf{L_A} = \mathbf{A}(\mathbf{A}+\sigma^2\mathbf{I})^{-1}$ and $\mathbf{L_B} =\mathbf{B}(\mathbf{B}+\sigma^2\mathbf{I})^{-1}$, then the following holds 
\begin{align*}
e_{\mathbf{B}} \leq e_{\mathbf{A}},
\end{align*}
where, $e_{\mathbf{A}}$ and $e_{\mathbf{B}}$ denoting the sum of the estimation error variance.
\end{lemma}
The proof is given in the Appendix. 

Applying these arguments and Lemma \ref{lem:CovMatMMSE}, we can ignore processing in the direction of frequency, as it does not yield enough gain for the amount of computational complexity spent on the NN processing.  

Now we focus only on the spatial processing. 
In a given cell, the number of different spatial channel covariance matrices could be assumed finite \cite{MMSE2018}. This assumption is not far from reality as in a typical cell, the channel covariance embeds UE's location information and UEs cannot be in any arbitrary location in a cell uniformly distributed. This already limits the search space of the feasible covariance matrices. In \cite{TurboAI_1}, the authors have already used this assumption to construct an algorithm that implicitly estimates the spatial covariance matrix of the input signal based on merely one single observation.

For the spatial sub-problem, we focus on learning only the weights in the spatial direction. The problem becomes 
\begin{align*}
\min_{\mathbf{U}} \|\mathbf{U}\mathbf{y}_s - \mathbf{h}_a\|_2^2 ,\;\; \forall s
\end{align*}
where, $\mathbf{y}_s = \big[y[s,0], y[s,1], \dots, y[s,M]\big]^H$ for any arbitrary $s$ and $\mathbf{U} \in \mathbb{C}^{M \times M}$ is the spatial weight matrix. The optimal solution is given as: 
\begin{align}\label{eq:mmse}
\mathbf{U} = \mathbf{R}_a (\mathbf{R}_a+ \sigma^2 \mathbf{I}_N )^{-1}
\end{align} 
where, $\mathbf{R}_a:= E[ \mathbf{h}_a \mathbf{h}_a^H] $ is the covariance matrix of the channel after the matched filter.

Inspired by the results from \cite{TurboAI_1, TurboAI_2}, we can learn the weights implicitly using a small neural network. The NN can estimate a coherently combined version of the input vector. We formulated the problem as a supervised learning using NN. The chosen architecture for the problem could be a simple three layers NN inspired by \cite{MMSE2018}. According to \cite{TurboAI_2}, this structure of the NN reduces the overall noise variance, by taking advantage of the spatial covariance among the entries of $\mathbf{y}_s$ at each subcarrier $s$. 

Similarities of the two problems of channel estimation and preamble detection becomes clear in the spatial domain. If we have an NN architecture and training mechanism, which is able to estimate the $\mathbf{h}_a$ from the noisy observations of $\mathbf{Y}$, we could apply the same structure to increase the SNR to have a "more" coherent combining of the values.

\begin{figure}
  \centering
    \includegraphics[width=0.49\textwidth]{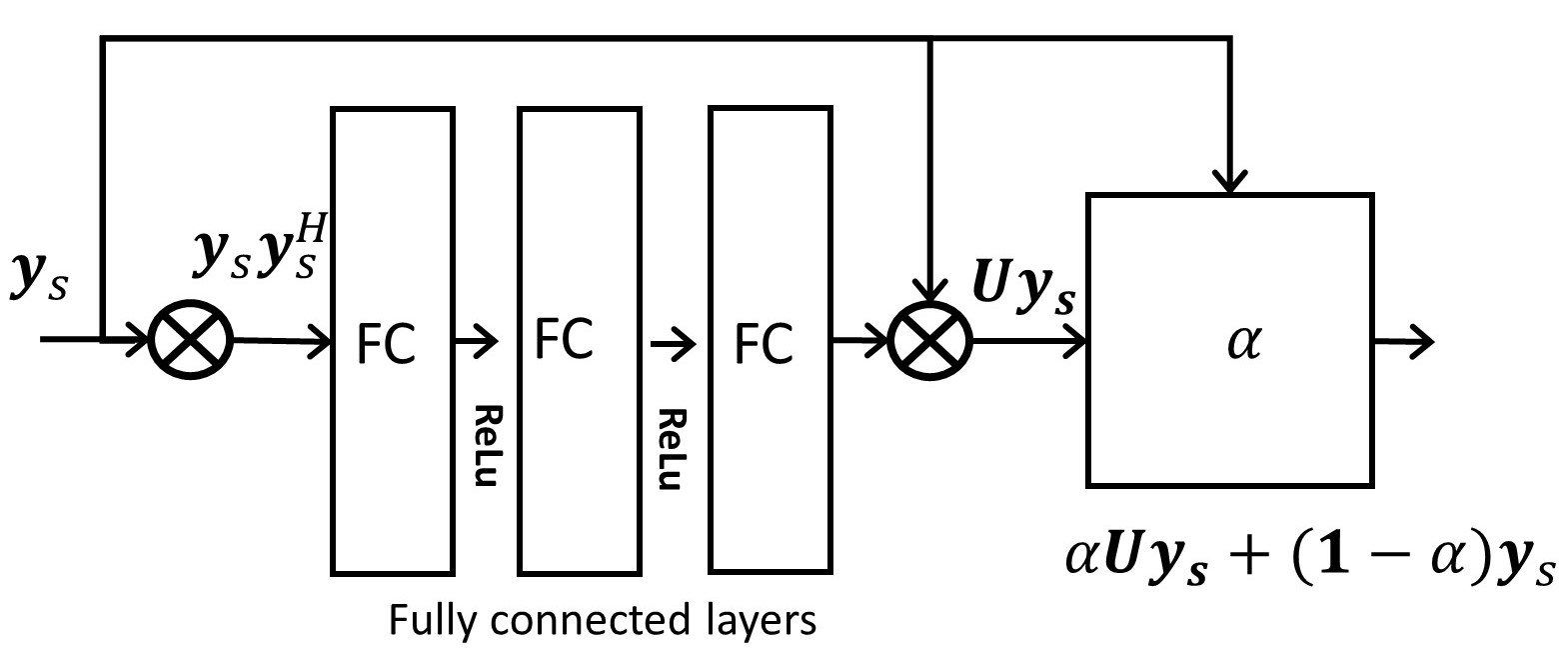}
    \caption{The neural network architecture with multiplicative layers and custom layer mitigate sensitivity to the SNR}
  \label{fig:block}
\end{figure}

\subsection{Extension to temporal direction}
\begin{figure}[t]
  \centering
    \includegraphics[width=0.48\textwidth]{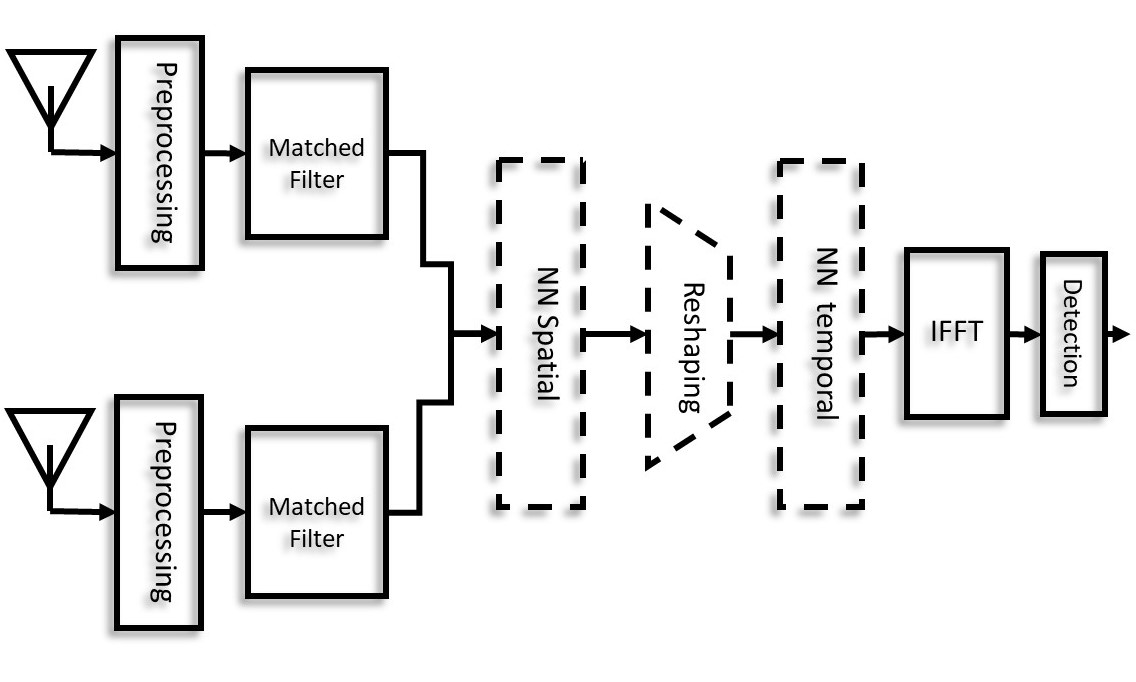}
    \caption{Full NN solution including spatial and temporal NN.}
    \label{fig:full}
\end{figure}

In standardized preamble construction to protect UEs at the cell edge, the same sequence is sent in repetition within a preamble. The repetition in the time domain adds one dimension to our analysis, very similar to the spatial dimension since at each given time, the same $z_u[n]$ is transmitted. The Kronecker approximation becomes: 
\begin{align*} 
E[\mathbf{GG}^{H}] \approx & E[\mathbf{h}_f \mathbf{Z}_{u,d,l} \mathbf{Z}^*_u (\mathbf{h}_f \mathbf{Z}_{u,d,l} \mathbf{Z}^*_u)^H] \otimes E[\mathbf{h}_a\mathbf{h}_a^H]] \\
 \otimes &E[\mathbf{h}_t\mathbf{h}_t^H]] 
\end{align*}
where, $\mathbf{h}_t \in \mathbb{C}^{T \times 1}$ is the temporal channel vector, and $T$ is the number of repeated sequences. 
\begin{align*}
\min_{\mathbf{Q}} \|\mathbf{Q}\mathbf{y}_{s,m} - \mathbf{h}_{t}\|_2^2 ,\;\; \forall s,m
\end{align*}
where, $\mathbf{y}_{s,m} = \big[y[0,s,m], y[1,s,m], \dots, y[T,s,m]\big]^H$ for any arbitrary $s,m$ and $\mathbf{Q} \in \mathbb{C}^{T \times T}$ is the temporal weight matrix, where T is number of repetitions in time domain. The optimal solution is given as: 
\begin{align}\label{eq:mmse}
\mathbf{Q} = \mathbf{R}_t (\mathbf{R}_t+ \sigma^2 \mathbf{I}_N )^{-1}
\end{align} 
where, $\mathbf{R}_t:= E[ \mathbf{h}_t \mathbf{h}_t^H] $ is the covariance matrix of the channel after the matched filter. The full solution comprising the spatial NN and temporal NN cascaded is depicted in a generic receiver chain in Fig.\ (\ref{fig:full}).

\subsection{Neural Network Architecture}

We use a simple three fully connected layers with rectified linear unit (ReLU) as nonlinear function between each two layers. The Input to the neural network admits one shot sample covariance matrix of the observed spatial observed signal slice as in: 
\begin{align*}
&\forall s\in \{0,\dots, S\} \text{ and } \forall t\in \{0,\dots, T\}:\\
&\textbf{Input:} \;\;\; \mathbf{y}_{s} \\
&\textbf{Label:} \;\;\; \mathbf{h_a}z_{u,l,d}[s]z^*_u[s]
\end{align*} 
For the temporal sub-problem, the same procedure applies into a different NN in parallel: 
\begin{align*}
&\forall s\in \{0,\dots, S\} \text{ and } \forall m\in \{0,\dots, M\}:\\
&\textbf{Input:} \;\;\; \mathbf{y}_{s,m} \\
&\textbf{Label:} \;\;\; \mathbf{h_t}z_{u,l,d}[s]z^*_u[s]
\end{align*} 

However, this method might underperform at the higher SNR inputs. 
The dependency of the NN-based detectors to input data SNR has been also observed by previous research on both preamble detection \cite{Chiara19} and channel estimation \cite{TurboAI_1}. We propose the use of a custom layer with trainable parameters to overcome this problem, specifically for the case of preamble detection by energy detector. The added layer proposes a weighted combination of the input (bypasser) and output, similar to the general concept of ResNets \cite{Resnet}. However, in our case the added layer has only one trainable parameter that could be potentially learned. This ensemble has been depicted in Fig.\ (\ref{fig:block}).We call this method Hybrid Neural network and Expert knowledge (HyNE). 

In a more practical scenario, we propose to train the first three layers of the NN separately, while using a heuristic design for $\alpha$. We propose deriving $\alpha$ from the instantaneous measured SNR of the received signal, assuming the knowledge of SNR.    
%
\section{Link Level Simulation Results}
The simulations have been carried out in a 3GPP compliant simulator. We selected the 3D UMa scenario according to 3GPP TR 38.901. The UEs are uniformly dropped within one tri-sectorized cell, which is about 8 km in size. The antenna configuration at UE is omni-directional with one antenna element. At BS sector antenna is directional with $65^{\circ}$ beamwidth in azimuth and elevation, 30 dB backward attenuation and $3^{\circ}$ mechanical downtilt. The BS has 8 antennas per sector in vertical arrangement with half a wavelength separation. The carrier frequency is 2 GHz. The UE have no mobility and PRACH power control is assumed as ideal in respect to target SNR. We carefully verified that the ideal power control assumption did not impact our key conclusions from our simulation studies. As preamble format, we have selected a short preamble format with sequence length 139 and 15 kHz subcarrier spacing. The resulting sequence duration is 66.67 microseconds. We did not employ sequence repetitions in our studies. In order to support cells with 8 km radius we used a cyclic prefix of 66.67 microseconds. The simulation tool had been cross validated in its PRACH detection performances for various scenarios in context of 3GPP 5G NR release 15 PRACH standardization study and work item. To demonstrate the generalizability of the HyNE detection concept, we trained the neural network with data from just a single BS sector and a single logical root sequence index. While in test phase or production mode, all possible (138) logical root indices and all 3 sectors of the BS contributed equally to the key performances metrics such as missed detection and false alarms probabilities and timing advance estimation errors. We further compared HyNE with conventional matched filter based method. Both in training sets and tests sets we kept the proportions of the two classes, i.e.\ signals vs noise equal, to avoid bias in the dataset.

\begin{figure}
  \centering
    \includegraphics[width=0.49\textwidth]{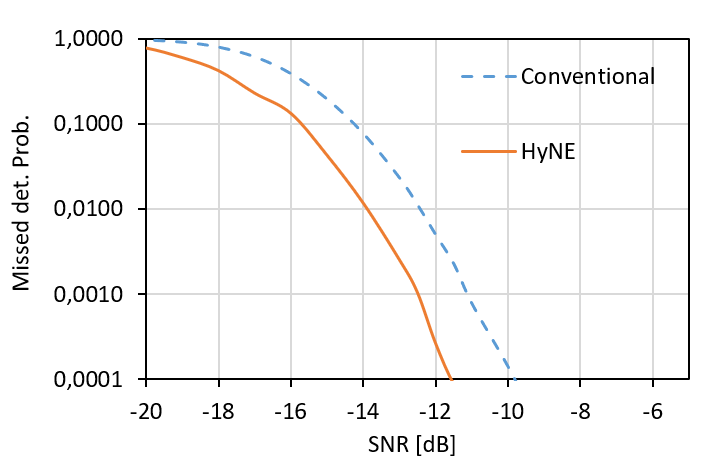}
    \caption{The probability of mis-detection for our proposed scheme (Hybrid) versus the conventional energy detector (Legacy).}
    \label{fig:misdetect}
\end{figure}

\begin{figure}
  \centering
    \includegraphics[width=0.49\textwidth]{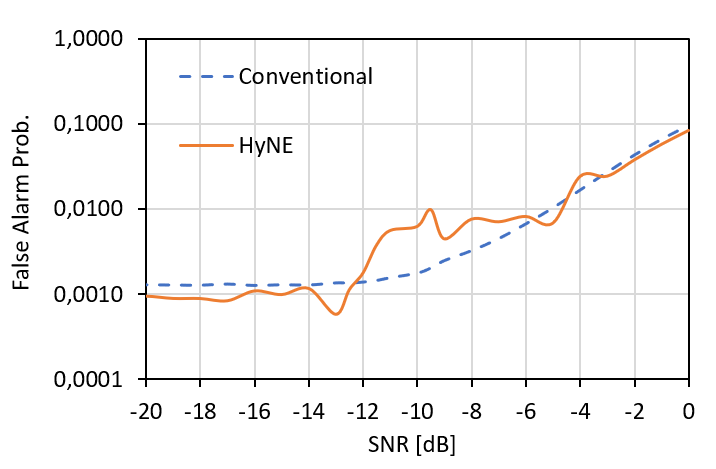}
    \caption{The probability of false alarm for our proposed scheme (Hybrid) versus the conventional energy detector (Legacy).}
    \label{fig:falseAlarm}
\end{figure}

In Fig.\ (\ref{fig:misdetect}) depicts the mis-detection probabilities for different SNRs of the UE. We observe clear benefits of using HyNE to the conventional approach. We have further compared the false-alarm probabilities for the sake fair comparison in Fig.\ (\ref{fig:falseAlarm}).  

\begin{figure}
  \centering
    \includegraphics[width=0.49\textwidth]{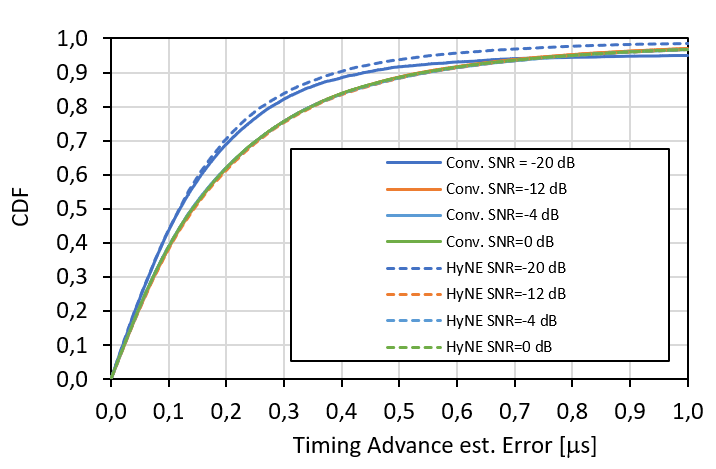}
    \caption{The CDF of HyNE and conventional methods are compared for different SNRs. The solid lines depict the performance of the conventional approach and the break lines demonstrate HyNE method.}
    \label{fig:TA}
\end{figure}

In Fig.\ (\ref{fig:TA}) we show the performance in estimating timing advanced. We plot the empirical cumulative distribution function (CDF) averaged over many channel realizations per SNR for each of the method. The solid lines show the conventional approach and the break lines show HyNE. The performance improvement for timing estimation is not significant, except for extremely small SNRs, i.e.\ below -20 dB, where HyNE can improve the results. 
%

\section{Conclusion}
We have investigated the problem of preamble detection and timing advanced estimation in terms of coherent combining of the signals from different antennas. We showed simple extension of this scheme for temporal direction as well. Our analysis proved that the frequency direction does not carry much cross correlation and therefore, might not contribute to the overall performance. This is due to the CAZAC properties of the Zadoff-Chu sequences. The goal of this research was to improve the current conventional scheme, while guaranteeing the performance for a wide range of channel models and SNRs. An interesting complimentary research could investigate dedicated learning of the NNs for a specific cell, benefiting from learning one environment. Furthermore, one can consider combining spatial and temporal NNs to improve HyNE even further.

\section{Appendix}

\noindent {\textit{Sketch of the Proof of Lemma \ref{lem:CovMatMMSE}}:} 

Before we begin with the proof with require the following lemma. This lemma could be seen as a result of \textit{Majorization theory} \cite{marshall11} and Gerschgorin circle theorem on spectral bounds, which we present without proof:
\begin{lemma} \label{lem:Gerschgorin}
If $\|\mathbf{A}\|_F \leq \|\mathbf{B}\|_F$ holds and $\textbf{diag}[\mathbf{A}] = \textbf{diag}[\mathbf{B}]$ then
\begin{align*}
\Lambda^\mathbf{A} \prec_w \Lambda^\mathbf{B}
\end{align*}
where, $\prec_w$ denotes weak Majorization and $\Lambda^\mathbf{A}$ and $ \Lambda^\mathbf{B} $ are vectors containing eigenvalues of $\mathbf{A}$ and $\mathbf{B}$ in sorted increasing order. Furthermore, $\textbf{diag}[\mathbf{A}]$ returns the diagonal elements of matrix $\mathbf{A}$. 
\end{lemma}
The estimation error variance is defined as: 
\begin{align*}
e_{\mathbf{A}} :=& \textbf{Tr}\big[E[(\hat{\mathbf{h}} - \mathbf{h})(\hat{\mathbf{h}} - \mathbf{h})^H]\big] \\
= & \textbf{Tr}\big[E[(\mathbf{L_A}\mathbf{y} - \mathbf{h})(\mathbf{L_A}\mathbf{y} - \mathbf{h})^H]\big] \\
= & \textbf{Tr}\big[E[(\mathbf{A}(\mathbf{A}+\sigma^2\mathbf{I})^{-1}\mathbf{y} - \mathbf{h})(\mathbf{A}(\mathbf{A}+\sigma^2\mathbf{I})^{-1}\mathbf{y} - \mathbf{h})^H]\big] \\
= & \textbf{Tr}\big[E[(\mathbf{A}(\mathbf{A}+\sigma^2\mathbf{I})^{-1}\mathbf{y}\mathbf{y}^H(\mathbf{A}+\sigma^2\mathbf{I})^{-1}\mathbf{A} \\
- & \mathbf{A}(\mathbf{A}+\sigma^2\mathbf{I})^{-1}\mathbf{y} \mathbf{h}^H 
- \mathbf{h}\mathbf{y}^H(\mathbf{A}+\sigma^2\mathbf{I})^{-1}\mathbf{A} 
+ \mathbf{h}\mathbf{h})^H]\big] \\
=& \textbf{Tr}\big[ \mathbf{L_A}\mathbf{A} - \mathbf{L_A} \mathbf{A} -\mathbf{L_A} \mathbf{A} + \mathbf{A}\big] 
\end{align*} 
which could be further simplified to the following: 
\begin{align} \nonumber
e_{\mathbf{A}} &= \textbf{Tr}[\mathbf{A} - \mathbf{L_A} \mathbf{A}] \\ \nonumber 
 &= \sum_i \frac{\sigma^2\lambda^\mathbf{A}_i}{\lambda^\mathbf{A}_i+\sigma^2} \\
 &=  \sum_i 
\frac{1}{\frac{1}{\sigma^2} + \frac{1}{\lambda^\mathbf{A}_i}} \label{eq:pr1}
\end{align}
where, we assume that $\mathbf{A = U} \textbf{Diag}[{\Lambda^\mathbf{A}}]\mathbf{U}^H$ and $\Lambda^\mathbf{A}: = [\lambda^\mathbf{A}_0, \dots, \lambda^\mathbf{A}_N]$ is vector of the eigenvalues of $\mathbf{A}$. 

From Lemma \ref{lem:Gerschgorin}, and results from \cite{marshall11}, we deduce: 
\begin{align} \nonumber
\Lambda^\mathbf{A} & \prec_w \Lambda^\mathbf{B} \Rightarrow \\ \nonumber
\frac{1}{\sigma^2\mathbf{1}_N} + \frac{1}{\Lambda^\mathbf{A}} & \succ_w \frac{1}{\sigma^2\mathbf{1}_N} + \frac{1}{\Lambda^\mathbf{B}} \Rightarrow\\
\frac{1}{\frac{1}{\sigma^2\mathbf{1}_N} + \frac{1}{\Lambda^\mathbf{A}}} & \prec_w \frac{1}{\frac{1}{\sigma^2\mathbf{1}_N} + \frac{1}{\Lambda^\mathbf{B}}} \label{eq: major1}
\end{align}
where, $\mathbf{1}_N \in \mathbb{N}^N$ is all ones vector. 
By definition of weak Majorization and (\ref{eq:pr1}) and (\ref{eq: major1}) yields: 
\begin{align}\label{eq: major2}
 e_{\mathbf{B}}=\sum_i \frac{\sigma^2\lambda^\mathbf{B}_i}{\lambda^\mathbf{B}_i+\sigma^2}   \leq    \sum_i \frac{\sigma^2\lambda^\mathbf{A}_i}{\lambda^\mathbf{A}_i+\sigma^2} =e_{\mathbf{A}} 
\end{align}


\section{Acknowledgment}
This work was partly funded by the German ministry of
education and research (BMBF) under grant 16KIS1184 (FunKI).

\bibliographystyle{IEEEtran}
\bibliography{Nokia}
\end{document}